\documentclass{iauc}
\usepackage{graphicx}
\usepackage{myaasmacros}
\input{epsf}

\def\ltsima{$\; \buildrel < \over \sim \;$}
\def\simlt{\lower.5ex\hbox{\ltsima}}   
\def\gtsima{$\; \buildrel > \over \sim \;$}
\def\simgt{\lower.5ex\hbox{\gtsima}}
\newcommand\bcite[1]{\citeauthor{#1} \citeyear{#1}}

\title[dSphs and the missing satellite problem] 
{The mass of dwarf spheroidal galaxies and the missing satellite problem}

\author[Read et. al.]{\small{J. I. Read$^1$\thanks{email:
  jir22@ast.cam.ac.uk} M. I. Wilkinson$^1$ N. Wyn
  Evans$^1$ G. Gilmore$^1$ \& Jan T. Kleyna$^2$}}

\affiliation{$^1$Institute of Astronomy, Cambridge University. 
  $^2$Institute for Astronomy, University of Hawaii}

\pubyear{2005}
\volume{198} 
\pagerange{119--126}
\date{?? and in revised form ??}
\jname{Near Field Cosmology with Dwarf Elliptical Galaxies}
\editors{Helmut Jerjen \& Bruno Binggeli}
\begin{document}

\maketitle

\begin{abstract}

We present the results from a suite of N-body simulations of the tidal
stripping of two-component dwarf galaxies comprising some stars and
dark matter. We show that recent kinematic data from the local group dwarf
spheroidal (dSph) galaxies suggests that dSph galaxies must
be sufficiently massive ($10^9 - 10^{10}$M$_\odot$) that tidal
stripping is of little importance for the stars. We discuss the
implications of these massive dSph galaxies for cosmology and galaxy
formation.


\end{abstract}

\firstsection 
\section{Introduction}\label{sec:introduction}
In the current cosmological paradigm all structure forms from the
successive mergers of smaller substructures
(see e.g. \bcite{1978MNRAS.183..341W}). Cosmological N-body 
simulations demonstrate that much of this substructure
will survive up to the present epoch in the form of dwarf galaxies
\citep{1996ApJ...462..563N}. While it is well known that cold dark
matter (CDM) theories over-produce this substructure, this
`sub-structure problem' can be solved by either suggesting that only
the most massive substructure halos form stars
\citep{2002MNRAS.335L..84S}; or by suppressing small
scale power (\bcite{2001ApJ...559..516A} and
\bcite{2002PhRvD..66d3003Z}). 

Either way, these solutions suggest that dSph galaxies in the local
group must be massive - some $\sim 10^9-10^{10}$M$_\odot$ in total
mass. At this mass, for all but
the most extreme plunging orbits, tidal stripping would not
affect the central stars and gas \citep{2003ApJ...584..541H}.

In contrast to this picture, there is compelling evidence to suggest
that tidal stripping could be extremely important for the formation and
evolution of the local group dSph
galaxies. \citet{2000Astro-Ph..0207467} and
\citet{2004ApJ...611L..21W} point out that all of the local group dSph
galaxies observed with deep enough photometry have a break in the
light profile which is similar to that observed in tidal stripping
simulations (see e.g. \bcite{1986ApJ...307...97A}, 
\bcite{2002AJ....124..127J} and
\bcite{2003ApJ...586L.123G}). Furthermore, \citet{2001ApJ...547L.123M} and
\citet{2001ApJ...559..754M} have suggested that the
very low internal angular momenta and spheroidal
morphologies observed in local group dSph galaxies could be explained
by tidal effects. However, to experience strong tidal effects, the
local group dSph galaxies must 
reside in low mass dark matter halos ($\simlt 10^9$M$_\odot$) which
appears to be at odds with naive cosmological predictions.

\citet{2004ApJ...608..663K} and \citet{2004ApJ...609..482K} have
recently argued that dSph galaxies might not inhabit the most massive
substructure halos. \citet{2004ApJ...609..482K} suggest that
reionisation sets an {\it epoch} at which the smallest collapsed
structures cannot
form stars, rather than having a mass scale below which star formation
ceases. This can go some way towards solving the mystery, but not all
of the way: even in the scheme proposed by \citet{2004ApJ...609..482K}
dSphs must inhabit the more massive halos on average. 

In order to resolve these issues we really require more
observational constraints from the local group dSph galaxies. Recently
\citet{2004ApJ...611L..21W} have found a sharp fall-off at $\sim 1$\,kpc
in the projected velocity dispersion of both the Draco and Ursa Minor
(UMi) dSph galaxies. At first glance, these data would seem to rule out
the tidal stripping model. It is well known that tidal stripping
produces heating of stars within the tidal radius (see
e.g. \bcite{1995ApJ...442..142O} and
\bcite{2003ApJ...586L.123G}). Thus the existence of a cold 
population at $\sim 1$\,kpc suggests that the tidal radius must lie
beyond this point and that dSph are then `high mass' galaxies.

In this paper we examine this argument more carefully. We present the
results from a suite of N-body numerical simulations of the tidal
stripping of two component dwarf galaxies comprising some stars and dark
matter, in orbit around a Milky Way-like galaxy.

\section{Results}\label{sec:results}

The initial conditions for the dwarf galaxies were set up as in 
\citet{2004ApJ...601...37K}, but for two-component spherical galaxies
comprising a dark matter halo and some stars. The satellite galaxies
were well resolved with $10^5$ star and $10^6$ dark matter particles.

\begin{figure} 
\begin{center}
  \epsfxsize=11.5truecm \epsfbox{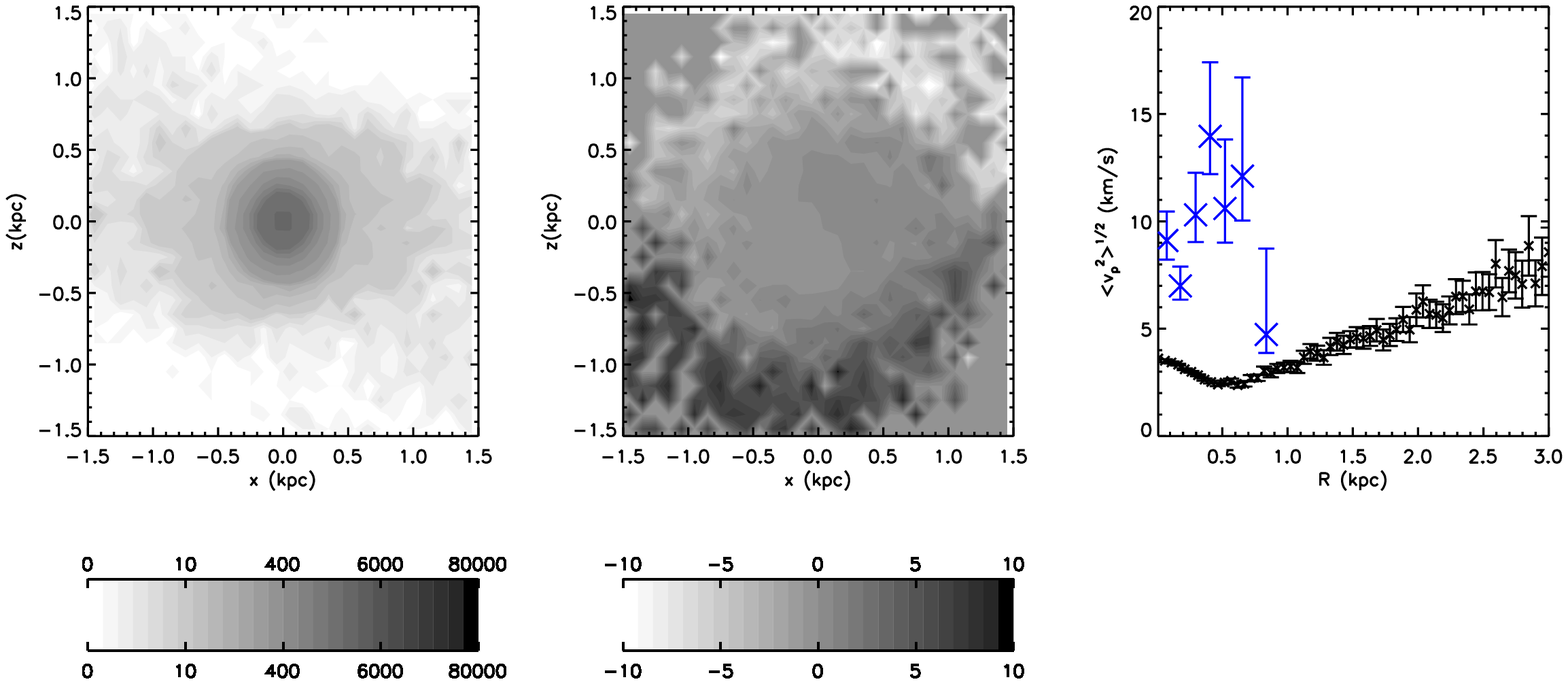}
  \epsfxsize=11.5truecm \epsfbox{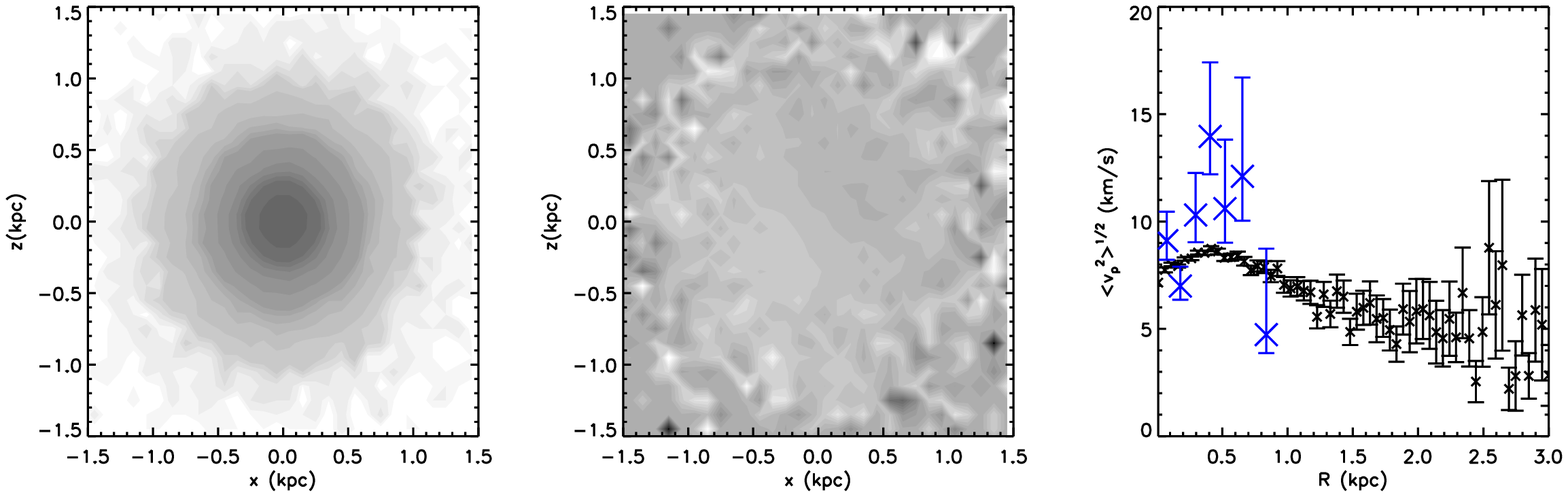}
  \epsfxsize=11.5truecm \epsfbox{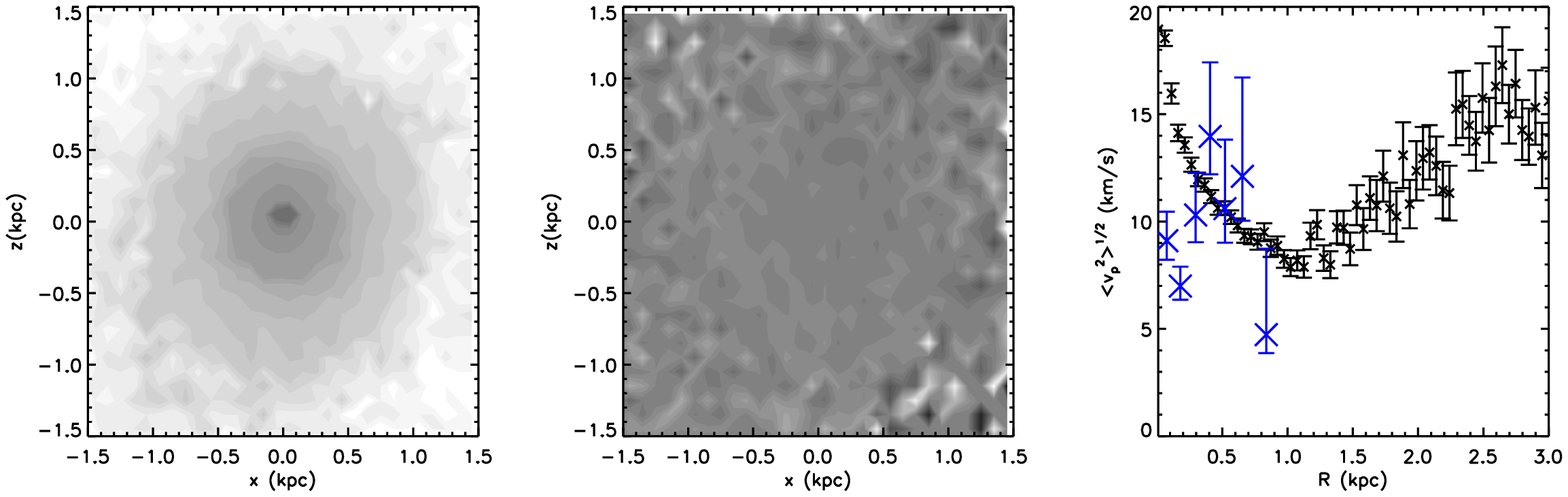}
\caption[]{Results from three models: A (top); B (middle) and C
  (bottom). Left panels show the projected surface density of the
  stars, the middle panels show the line of sight velocities of the
  stars and the right panels show the projected velocity dispersion of
  the stars as a function of radius. The contour bars beneath the top
  left and middle panels show the number of star particles and the
  projected velocity in km/s respectively. Over-plotted on the right panels
  (large crosses) are the data points from the Draco dSph galaxy
  \citep{2004ApJ...611L..21W}.}
\label{read.fig1} 
\end{center}
\vspace{-4.5mm}
\end{figure}

The N-body integration for the orbit of the dwarf galaxy about the
Milky Way was then performed using a modified version of
the GADGET N-body code \citep{2001NewA....6...79S}. The
code allowed for a fixed potential to be introduced to model the Milky
Way. We used a Miyamoto-Nagai potential for the Milky Way disc and
bulge and a logarithmic potential for the Milky Way dark matter
halo. The simulations, initial conditions and model parameters are
described in more detail in \citet{Readinprep1}.

Figure \ref{read.fig1} shows the results for three very different
dwarf galaxies on orbits around the Milky Way after $\sim
10$\,Gyrs. In each case the orbiting galaxy is currently at apocentre.

The first model (model A) is one in which the satellite galaxy
contains {\it no dark matter}. It was a Plummer sphere
\citep{1987gady.book.....B} with total mass and
scale length of $10^7$M$_\odot$ and 0.23\,kpc. The galaxy was placed on
a plunging orbit
with apocentre and pericentre of 80 and 20\,kpc respectively. Notice, as
published by previous authors, 
that the isodensity contours (left panel) show the tidal `S-shape'
caused by the leading and trailing tidal debris. Several authors have
argued that there is evidence for such shapes in the isodensity
contours of the local group dSph galaxies (see
e.g. \bcite{2003ApJ...586L.123G} and \bcite{Munoz:2005be}); and that
this, therefore, implies that the local group dSph galaxies must be
strongly tidally distorted. Consider, however, the middle panel for
the line of sight velocities. There is a strong velocity gradient
across the the galaxy which is not observed in Draco or Ursa Minor
(UMi) \citep{2004ApJ...611L..21W}. Furthermore, the projected velocity
dispersion is rising beyond the tidal radius, whereas in both Draco
and Ursa Minor it is observed to drop sharply
\citep{2004ApJ...611L..21W}. Projections which give
distorted isodensity contours, small velocity gradients and near-flat
projected velocity dispersions are very rare \citep{Readinprep1}. In
{\it no} projection was the projected velocity dispersion of the stars
found to fall.

Even if we are viewing the local group dSph galaxies from a
projection which minimises any kinematic trace of tidal stripping,
notice that the central velocity dispersion in model A is much smaller
than that observed in the Draco dSph
galaxy (see data points with large crosses). As shown by
\citet{2001ApJ...563L.115K}, 
the only way to reproduce such a large central velocity dispersion
with the stars we see in Draco is to introduce a significant dark
matter component (and see also \bcite{2000Astro-Ph..0302287}).

Model B contained a significant Plummer dark matter component with mass $14
\times 10^7$M$_\odot$ and scale length 0.5\,kpc; its orbit was identical
to model A. Even after 10\,Gyrs the galaxy has regular isodensity
contours, no velocity gradient and a falling 
projected velocity dispersion. This is because the initial mass and
concentration of dark matter was chosen to be sufficiently high that
tidal stripping produced almost no effect on the stars over
10\,Gyrs. Notice that for this model, the central velocity dispersion is
much closer to that of Draco. 

Finally, model C was chosen to have a dark matter halo consistent with
those observed in cosmological simulations. We used a Hernquist
profile with mass $10^{10}$M$_\odot$ and scale length $1.956$\,kpc. In
order for it to have a final velocity dispersion low enough given this
starting mass we had to choose an extreme orbit with a pericentre of just
6\,kpc\footnote{The central velocity dispersion is much lower than in the initial
conditions for this model due to disc shocking and tidal stripping
(see \bcite{Readinprep1} for more details). It does tend to zero at
the centre as it should, but this is disguised somewhat by the choice
of bin size.}. At this radius
the galaxy would likely affect the Milky Way 
disc (and dynamical friction would become important) and so already
the orbit is quite unrealistic. However, notice that even for such an
extreme orbit, the central velocity dispersion is still too large
after 10\,Gyrs to be consistent with Draco.

\vspace{-3mm}
\section{Conclusions}

In conclusion, tidal stripping cannot be very strong for many, if not
all, of the local group dSphs. Strong tidal stripping, which would produce
distorted isodensity contours, also leads to velocity gradients and
flat or rising projected velocity dispersions - neither of which are
observed in the local group dSphs for which we have good kinematic
data (but see also \bcite{Munoz:2005be}). This suggests that
dSph galaxies must be sufficiently massive such that tidal stripping
is of little importance for the stars. Either they are on orbits with large
pericentres, in which case they can have masses as low as $\sim
10^8$M$_\odot$ \citep{2001ApJ...563L.115K}; or they are on more
extreme orbits in which case they must be $\sim 10^9-10^{10}$M$_\odot$
depending on the extremity of the orbit. Our current cosmological
paradigm would favour the latter hypothesis, but this leaves us with a
puzzle: if the dSph are really as massive as $\sim 10^{10}$M$_\odot$
and have dark matter densities which are cosmologically consistent
then they would have central velocity dispersions which are too large 
to be consistent with Draco or UMi - even after significant tidal
stripping and shocking.

\bibliographystyle{mn2e}
\bibliography{/home/jir22/More_space__/LaTeX/BibTeX/refs}

\end{document}